\documentclass{article}

\usepackage{amssymb}
\usepackage{spconf,amsmath,graphicx}

\usepackage{float}
\usepackage{tikz}
\usepackage{pgfplots}
\usepackage{multirow}
\usepackage{bigints}
\usepackage[shortlabels]{enumitem}
\usepackage{empheq} 
\usepackage{cases}
\usepackage{caption}

\usepackage{array}
\usepackage{dblfloatfix}    
\usetikzlibrary{shapes,arrows,positioning,calc}
\definecolor{alizarin}{rgb}{0.82, 0.1, 0.26}
\definecolor{bleudefrance}{rgb}{0.19, 0.55, 0.91}
\definecolor{britishracinggreen}{rgb}{0.0, 0.26, 0.15}
\definecolor{internationalkleinblue}{rgb}{0.0, 0.18, 0.65}
\definecolor{tangerine}{rgb}{0.95, 0.52, 0.0}
\definecolor{cornellred}{rgb}{0.7, 0.11, 0.11}
\definecolor{dl_inter_blue}{rgb}{0.2, 0.2, 1.0}
\definecolor{dl_inter_red}{rgb}{1.0, 0.2, 0.2}
\definecolor{amethyst}{rgb}{0.6, 0.4, 0.8}
\definecolor{arsenic}{rgb}{0.23, 0.27, 0.29}
\definecolor{amber}{rgb}{1.0, 0.75, 0.0}
\definecolor{applegreen}{rgb}{0.55, 0.71, 0.0}
\definecolor{midnightblue}{rgb}{0.1, 0.1, 0.44}
\definecolor{emerald}{rgb}{0.31, 0.78, 0.47}

\title{Binary Probability Model for Learning based Image Compression}
\name{Th\'{e}o LADUNE\textsuperscript{*$\dagger$}, Pierrick PHILIPPE\textsuperscript{*}, Wassim HAMIDOUCHE\textsuperscript{$\dagger$}, Lu ZHANG\textsuperscript{$\dagger$}, Olivier D\'{E}FORGES\textsuperscript{$\dagger$}}
\address{\textsuperscript{*}Orange Labs, 4 rue du Clos Courtel, 35512,
Cesson-S\'{e}vign\'{e}, France\\
\normalsize{\texttt{firstname.lastname@orange.com}}\\
\textsuperscript{$\dagger$}Univ. Rennes, INSA Rennes,
CNRS, IETR --  UMR 6164, Rennes, France\\ 
\normalsize{\texttt{firstname.lastname@insa-rennes.fr}}
}

\pdfoutput=1
\begin{document}
%
\maketitle

\begin{abstract}
  In this paper, we propose to enhance learned image compression systems with a
  richer probability model for the latent variables. Previous works model the
  latents with a Gaussian or a Laplace distribution. Inspired by binary
  arithmetic coding, we propose to signal the latents with three binary values and one
  integer, with different probability models.
    
  A relaxation method is designed to perform gradient-based training. The richer
  probability model results in a better entropy coding leading to lower rate.
  Experiments under the Challenge on Learned Image Compression (CLIC) test
  conditions demonstrate that this method achieves 18~\% rate saving compared to
  Gaussian or Laplace models.
\end{abstract}
\section{Introduction} 
\label{sec:intro}

Data compression can be summarized in three main steps. First, the input signal
is encoded into more compact variables called latents. Then, the latents are
transmitted with a coding method achieving a rate near to the Shannon entropy.
Lastly, the input signal is decoded from the latents.
As a real number has an infinite information quantity (\textit{i.e.} an infinite
number of bits), lossy coding methods only work with finite set of values. To
address this issue, latents are quantized, introducing distortion on both the
latents and the reconstructed signal. 

Lossy image compression can thus be expressed as an optimization problem:
jointly minimizing the distortion and the rate (\textit{i.e.} information
in the latents). Traditional coding approaches such as JPEG or BPG
(HEVC-based image
compression)~\cite{Wallace:1991:JSP:103085.103089,BPG_web_page} typically solve
this problem using linear predictions and transforms. Deep neural networks can 
learn complex non-linear functions, making them well suited to
reach better optimum and coding efficiency. However, the discrete nature of the
data sent from the encoder to the decoder makes the objective function
non-differentiable and prevents optimizing end-to-end systems with gradient-based methods. 

In~\cite{DBLP:conf/iclr/BalleLS17}, authors suggest to replace quantization with
additive noise and propose an interpolation of the rate function.  A different
quantization approximation is presented in \cite{DBLP:conf/iclr/TheisSCH17}.
These works show promising results, outperforming the JPEG
standard.

Entropy coding requires an estimate of the latents
probability density function (PDF). Whereas previous works use a fixed-PDF
model, Ball\'{e} { \it et al.} introduce hyperpriors in
\cite{DBLP:conf/iclr/BalleMSHJ18}, consisting in side-information
conditioning each latent PDF. This more accurate probability model brings
important performance gains. Minnen {\it et al.} and Lee {\it et al.}
\cite{DBLP:conf/nips/MinnenBT18,DBLP:conf/iclr/LeeCB19} add an autoregressive
model (ARM) to infer PDF parameters from previously sent values. However, such
systems lead to a prohibitive decoding time due to the sequential nature of
the ARM which is not suited for GPU processing.

In 2019, the Challenge on Learned Image Compression (CLIC) \cite{CLIC_web_page}
was held at the Conference on Computer Vision and Pattern Recognition (CVPR),
providing a common evaluation framework to the learned image compression
community. Proposed end-to-end systems \cite{Zhou_2019_CVPR_Workshops,
Wen_2019_CVPR_Workshops} composed of a hyperprior and an ARM
outperformed BPG \cite{BPG_web_page}.

Improvements of the latents probability model are the main reason behind the
successive performance gains. In this paper, we propose a more accurate
estimate of the latents PDF widely inspired by the HEVC binarization
process~\cite{Sullivan:2012:OHE:2709080.2709221}. Based upon Minnen's
work~\cite{DBLP:conf/nips/MinnenBT18}, we present a new relaxation method for a
discrete rate function. This allows to leverage the richer probability
model providing either better performance with the same complexity or similar
performance with a lightweight coding system.

\section{Proposed Method}

\subsection{Framework description}

\label{sec:method}
\begin{figure}
  \centering
  \includegraphics[scale=0.30]{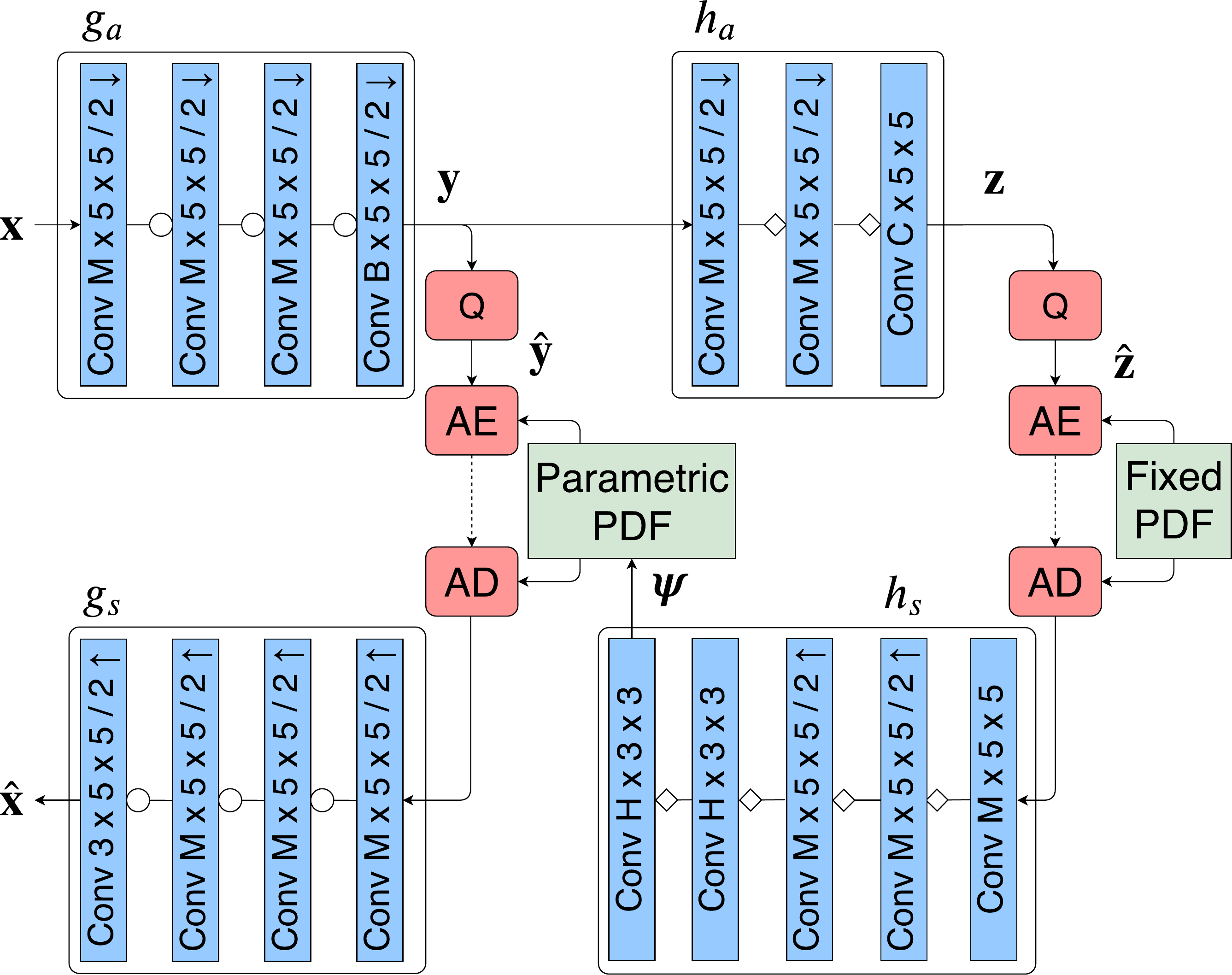}
  \caption{Network architecture. Rounded arrows denote GDN \cite{DBLP:conf/iclr/BalleLS17} and squared arrows LeakyReLU. 
  Convolution parameters are: filters number $\times$ kernel height $\times$ width / stride.
  Upscaling convolutions are transposed convolutions.}
  \label{architecture}
\end{figure}

The work carried out in this paper is based upon Ball\'{e}
and
Minnen's work \cite{DBLP:conf/iclr/BalleLS17,DBLP:conf/iclr/BalleMSHJ18,DBLP:conf/nips/MinnenBT18}.
Their framework for training end-to-end lossy compression system is
explained in this section. The architecture is the one described
in~\cite{DBLP:conf/nips/MinnenBT18}. Fig.~\ref{architecture} illustrates the
coding scheme which can be summarized as:
\begin{enumerate}
  \itemsep0em 
  \item Encoding the input image $\mathbf{x}$ into latents
  $\mathbf{y} = g_a(\mathbf{x}; \boldsymbol{\theta}_e)$;
  \item Encoding the hyperprior $\mathbf{z} = h_a(\mathbf{y}; \boldsymbol{\theta}_{he})$;
  \item Quantizing $\hat{\mathbf{z}} = Q(\mathbf{z})$, $\hat{\mathbf{y}} =
  Q(\mathbf{y})$ with a unitary uniform scalar quantizer;
  \item Lossless arithmetic encoding (AE) and decoding (AD);
  \item Decoding PDF parameters $\boldsymbol{\psi} =
  h_s(\hat{\mathbf{z}}; \boldsymbol{\theta}_{hd})$;
  \item Decoding $\hat{\mathbf{y}}$ to reconstruct the input image $\hat{\mathbf{x}}
  = g_s(\hat{\mathbf{y}}; \boldsymbol{\theta}_d)$.
\end{enumerate}

The set of neural network parameters $\left\{\boldsymbol{\theta}_e,\boldsymbol{\theta}_d,\boldsymbol{\theta}_{he},\boldsymbol{\theta}_{hd}\right\}$
is learnt by minimizing a rate-distortion trade-off
\begin{equation*}
  \mathcal{L}(\lambda) = D(\mathbf{x}, \hat{\mathbf{x}}) + \lambda \left  (R(\hat{\mathbf{y}}) + R(\hat{\mathbf{z}}) \right ).
\end{equation*}

In this work, the distortion is computed through the mean-squared error
$D(\mathbf{x},\ \hat{\mathbf{x}}) = \mathbb{E}_{\mathbf{x} \sim
p_{\mathbf{x}}}\left [ || \mathbf{x} - \hat{\mathbf{x}}||^2 \right ]$.

Latents $\hat{\mathbf{y}}$ and the hyperprior $\hat{\mathbf{z}}$ are encoded
with arithmetic coding, a lossless coding method achieving a rate near to
Shannon entropy
\begin{equation*}
  R(\hat{\mathbf{y}}) = 
  \mathbb{E}_{\hat{\mathbf{y}} \sim m} [L(\hat{\mathbf{y}};  \mathbb{P}_{\hat{\mathbf{y}}})]=
  \mathbb{E}_{\hat{\mathbf{y}} \sim m} [-\log_2  \mathbb{P}_{\hat{\mathbf{y}}}(\hat{\mathbf{y}})],
\end{equation*}
where $m$ denotes the distribution of latents (which is unknown) and $L$ is the code
length computed thanks to the probability model $P_{\hat{\mathbf{y}}}$.
This can be re-written as \cite{DBLP:conf/iclr/LeeCB19}:
\begin{equation*}
  R(\hat{\mathbf{y}}) =  H(m) + D_{KL}(m\ ||\ \mathbb{P}_{\hat{\mathbf{y}}}),
\end{equation*}
where $D_{KL}$ denotes the Kullback-Leibler divergence. Thus, minimizing the
rate implies to jointly lower the entropy $ H(m)$ of $\hat{\mathbf{y}}$ and
properly match the distribution $m$ with the probability model
$\mathbb{P}_{\hat{\mathbf{y}}}$. This also holds for rate of $\hat{\mathbf{z}}$.

Training neural networks relies on gradient-based algorithms, requiring all
operations to be differentiable. Because quantization derivative is null almost
everywhere, it is modeled as an additive uniform noise during training
\cite{DBLP:conf/iclr/BalleLS17}
\begin{equation*}
  \tilde{\mathbf{y}} = \mathbf{y + u} \Rightarrow p_{\tilde{\mathbf{y}}} = p_{\mathbf{y}} * p_{\mathbf{u}},\ \mathbf{u} \sim \mathcal{U}( -\tfrac{1}{2},\ \tfrac{1}{2} ),
\end{equation*}
where $p$ denotes probability distribution. Continuous interpolation $\tilde{L}(\tilde{\mathbf{y}};
p_{\tilde{\mathbf{y}}}) = -\log_2 p_{\tilde{\mathbf{y}}}(\tilde{\mathbf{y}})$ of the code length function is used as a proxy to optimize discrete
$L(\hat{\mathbf{y}}; P_{\hat{\mathbf{y}}})$. The same goes for $\hat{\mathbf{z}}$
and the loss function becomes
\begin{equation}
  \mathcal{L}(\lambda) = \mathbb{E}_{\mathbf{x} \sim p_{\mathbf{x}}}\left[||\mathbf{x} - \hat{\mathbf{x}}||^2 +
  \lambda (\tilde{L}(\tilde{\mathbf{y}}; p_{\tilde{\mathbf{y}}}) 
  + \tilde{L}(\tilde{\mathbf{z}}; p_{\tilde{\mathbf{z}}})\right].
  \label{loss_balle}
\end{equation}

The hyperprior distribution $p_{\tilde{\mathbf{z}}}$ is estimated through a fixed model described
in \cite{DBLP:conf/iclr/BalleMSHJ18}. Each latent $y_i$ is coded independently and their distribution $p_{y_i} \sim \mathcal{N} \left ( \mu_i, \sigma_i \right )$ is decoded from the hyperprior
\begin{align}
  \tilde{L}(\tilde{\mathbf{y}}; p_{\tilde{\mathbf{y}}}) &= \sum_i \tilde{L}(\tilde{y}_i, p_{\tilde{y}_i}) = \sum_i -\log_2 \left(p_{y_i} * p_u \right) (\tilde{y}_i) \nonumber \\
  &= \sum_i -\log_2 \int_{\tilde{y}_i - \frac{1}{2}}^{\tilde{y}_i + \frac{1}{2}} \mathcal{N}(u;\ \mu_i, \sigma_i)\ \mathrm{d}u.\label{balle_relaxed_rate}
\end{align}

In this paper, we enhance the probability model $p_{y_i}$ in order to improve
the entropy coding efficiency. As in traditional video coding, latents are
transmitted in a binary version, allowing a more accurate model $p_{y_i}$.

\subsection{Binary probability model}

For the sake of clarity, latents index is omitted \textit{i.e.} $y$ stands for
any $y_i$. The purpose of this work is to relax assumptions on $p_y$. To do so,
each latent is represented with three binary values and one integer with
separate probability model. First, the expectation $\mu$ is decoded from the
hyperprior and used to center $y$ before quantization: $\hat{y} = Q(y - \mu)$.
Each $\hat{y}$ is then signaled as described in Table~\ref{flags}.

\begin{table}[h]
  \centering
  \begin{tabular}{c|cccc|c}
    \multirow{2}{*}{$\hat{y}$} &  \multicolumn{4}{c|}{Elements transmitted} & \multirow{2}{*}{Code length $L_{bin}$}\\
                               & $G_0$ & $G_1$ & $S$      & $E$          & \\
    \hline
    $0$                        & $0$   &       &          &               & $L_{G_0}$\\
    $\pm\ 1$                   & $1$   & $0$   & $\pm 1$  &               & $L_{G_0} + L_{G_1} + L_S$\\
    $\pm\ k$                   & $1$   & $1$   & $\pm 1$  & $k$           & $L_{G_0} + L_{G_1} + L_S + L_E$\\
  \end{tabular}
  \caption{$G_0$ (respectively $G_1$) stands for greater than zero (respectively one), $S$ for sign and $E$ for explicit.}
  \label{flags}
\end{table}

Flags $G_0$ and $G_1$ are transmitted using an entropy coding method, their code
length is estimated as
\begin{equation*}
  L_{G_X} = \left\{\begin{array}{ll}
    -\log_2 P_{G_X} &\text{ if } G_X = 1,\\
    -\log_2 \left(1 - P_{G_X}\right) &\text{ otherwise}
  \end{array}  
  \right. X = \left\{0,\ 1\right\}.
\end{equation*}
Probabilities $P_{G_0}$ and $P_{G_1}$ are decoded from the hyperprior
$\hat{\mathbf{z}}$. The sign flag is assumed equiprobable costing $L_S~=~1$ bit.
A latent $|\hat{y}| \geq 2$ is explicitly transmitted with a code length estimated as
\begin{equation}
  L_{E}(k) = - \log_2 P_{\hat{y}}(|\hat{y}| = k\ \big| \ |\hat{y}| > 1).
  \label{remaining}
\end{equation}
Here, $p_y$ is modelled as a centered Laplace distribution with $\sigma$
decoded from the hyperprior. Equation \eqref{remaining} becomes
\begin{equation}
  L_{E}(k) = -\log_2 \left( \frac{2 \int_{k-0.5}^{k+0.5}\mathcal{L}(u;0, \sigma)\ \mathrm{d}u}{1 - \int_{-1.5}^{1.5}\mathcal{L}(u;0, \sigma)\ \mathrm{d}u}\right).
  \label{explicit}
\end{equation}

The total code length $L_{bin}$ is obtained by adding up all transmitted
elements (\textit{cf.} Table \ref{flags}). All $\hat{y} \in \left\{-1,\ 0,\
1\right\}$ are no longer constrained to a pre-determined distribution as $P_{\hat{y}}$ can represent any symmetrical probability distribution in this interval.
The entropy coding of each latent $y$ requires the set $\left\{\mu, \sigma, P_{G_{0}},
P_{G_{1}}\right\}$. Hence, the decoded hyperprior
$\boldsymbol{\psi}$ has four features per
$\hat{y}$: in Fig. \ref{architecture} $H = 4B$.

\subsection{Relaxed rate}

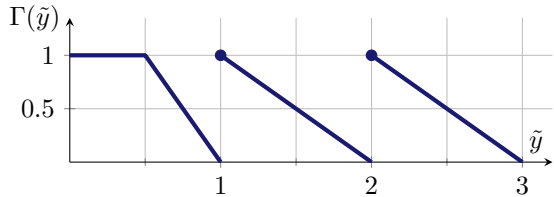
\begin{figure}
\centering
\pgfplotsset{compat=1.12}
\pgfplotsset{soldot/.style={color=midnightblue,only marks,mark=*}} \pgfplotsset{holdot/.style={color=midnightblue,fill=white,only marks,mark=*}}
\begin{tikzpicture}
  \begin{axis}[
    height=3.5cm,
    width=8cm,
    axis lines = middle,
    xlabel=$\tilde{y}$,
    ylabel=$\Gamma(\tilde{y})$,
    xtick distance   = 1,
    minor x tick num = 1,
    ytick distance   = 0.5,
    minor y tick num = 0,
    ylabel style={left},
    grid             = both,
    ymin=0, ymax=1.35, xmin=0, xmax=3.2,
    samples=200]
    \addplot[midnightblue, ultra thick, domain=0:1] {x > 0.5 ? 2 - 2 * x: 1};
    \addplot[midnightblue, ultra thick, domain=1:2] {floor(x) + 1 - x};
    \addplot[midnightblue, ultra thick, domain=2:3] {floor(x) + 1 - x};

    \addplot[soldot] coordinates{(1,1)(2,1)};

  \end{axis}
  \end{tikzpicture}
  \caption{The weighting function $\Gamma$.}
  \label{gamma_func}
\end{figure}

The previous section proposes a richer representation of $P_{\hat{y}}$.
During training, discrete $\hat{\mathbf{y}}$ is replaced by a continuous
$\tilde{\mathbf{y}}$, requiring the interpolation of the code length function
$\tilde{L}$. As no hypothesis is made on $p_y$, eq. \eqref{balle_relaxed_rate}
can not be used directly. A new interpolation $\tilde{L}_{bin}$ is introduced as
a weighted sum of the two nearest integer rates:
\begin{equation*}
  \tilde{L}_{bin}(\tilde{y}) = \Gamma(|\tilde{y}|) L_{bin}(\lfloor \tilde{y} \rfloor) + \left(1 - \Gamma(|\tilde{y}|)\right) L_{bin}(\lfloor \tilde{y} \rfloor + 1),
\end{equation*}
where $\lfloor \cdot \rfloor$ denotes the floor function. $\Gamma(\tilde{y})$ is
a weighting function defined with linear segments and depicted in Fig.
\ref{gamma_func}.
The main design constraint on the weighting function $\Gamma$ is to ensure that
$\tilde{L}_{bin}(k) = L_{bin}(k)$ for all integers $k$ to make training and
inference metrics coherent. Because sending $\hat{y} = 0$ requires only one
element ($G_0$), the optimization process results in zeros being the
most present value. The flat zone in $[0, \tfrac{1}{2}]$ is used to make the
optimization focus more on the cost of zeros. In $[1, +\infty]$ interval, $\Gamma$ is a
simple linear weighting based on the distance to the nearest integer.
With the relaxed rate, the loss function becomes:
\begin{equation*}
  \mathcal{L}(\lambda) = \mathbb{E}_{\mathbf{x} \sim p_{\mathbf{x}}}[||\mathbf{x} - \hat{\mathbf{x}}||^2 -
  \lambda (\tilde{L}_{bin}(\tilde{\mathbf{y}}) +\tilde{L}(\tilde{\mathbf{z}}; p_{\tilde{\mathbf{z}}}))].
\end{equation*}

\section{Experimental Results}

\subsection{Performance on CLIC low-rate task}

\begin{figure*}[htp]
  \begin{minipage}[l]{1.0\columnwidth}
      \centering
      \begin{tabular}{cc||cc|cc}
        \multirow{3}{*}{Systems} & \multirow{3}{*}{$M$} &  \multicolumn{2}{c|}{Validation} & \multicolumn{2}{c}{Test}\\
                                 &                      & PSNR       & BD rate &   PSNR & BD rate\\
                                 &                      & [dB]       & [\%]    &   [dB] & [\%]\\
                                 \hline
        JPEG & \multirow{2}{*}{/} & 26.31 & \multirow{2}{*}{/} & 25.10 & \multirow{2}{*}{/}\\
        BPG & & 30.84 &  & 29.60 & \\
        \hline
        Gaussian & \multirow{3}{*}{64} & 30.10 & Ref. & 28.87 & Ref.\\
        Laplacian &  & 30.22 & -5.9& 28.99 & -7.5\\
        Binary &  & \textbf{30.48} & \textbf{-14.4}& \textbf{29.26} & \textbf{-18.3} \\
        \hline
        Gaussian & \multirow{3}{*}{192} & 30.56 & Ref.& 29.31 &Ref. \\
        Laplacian & & 30.51 & 2.1 & 29.26 & 3.1 \\
        Binary & & \textbf{30.68} & \textbf{-7.5} & \textbf{29.49} & \textbf{-9.1}\\
        \hline
      \end{tabular}
  \caption{Latents probability models performances on CLIC validation and test sets. PSNR are given at 0.15~bpp. BD rates are computed with the Gaussian system as reference.}
  \label{comprehensive_results}
    \end{minipage}
  \hfill{}
  \begin{minipage}[r]{1.0\columnwidth}
      \centering
      \begin{tikzpicture}
        \begin{axis}[
        height=5cm,
        width=6.5cm,
        ylabel=PSNR (dB),
        xlabel=Number of convolution features $M$,
        ymin=28.50, ymax=30.00,
        ytick distance=0.5,
        yminorgrids=true,
        ymajorgrids=true,
        minor y tick num=4,
        ybar, bar width=15,
        enlarge x limits=0.55,
        symbolic x coords={0,64,192, 256},
        xtick=data,
        legend entries={Gaussian,
                        Laplacian,
                        Binary},
        legend style={at={(1.365,0.5)},anchor=east},
        title={$PSNR$ at 0.15 bpp -- CLIC 2019 test set}
        ]
          \draw[ultra thick, dashed, color=black] (axis cs:0,29.60) -- node[above] {BPG} (axis cs:256,29.60);
          \addplot[ybar, fill=midnightblue] coordinates {(64,28.87)(192,29.31)};
          \addplot[ybar, fill=emerald] coordinates {(64,28.99)(192,29.26)};
          \addplot[ybar, fill=cornellred] coordinates {(64,29.26)(192,29.49)};
        \end{axis}
        \end{tikzpicture}
    \caption{Latents probability models performances.}
    \label{res_test}
    \end{minipage}
\end{figure*}

\newcommand{\mysize}{0.28} 
\newcommand{\picsize}{0.11} 
\newcommand{\ftsize}{0.30} 
\begin{figure*}[!b]
  \begin{center}
  \setlength\tabcolsep{1pt} 
  \begin{tabular}{cccccccccccc}
    \multicolumn{6}{c}{Input image $\mathbf{x}$} & \multicolumn{6}{c}{Decoded image $\hat{\mathbf{x}}$}\\
    \multicolumn{6}{c}{\includegraphics[scale=\picsize]{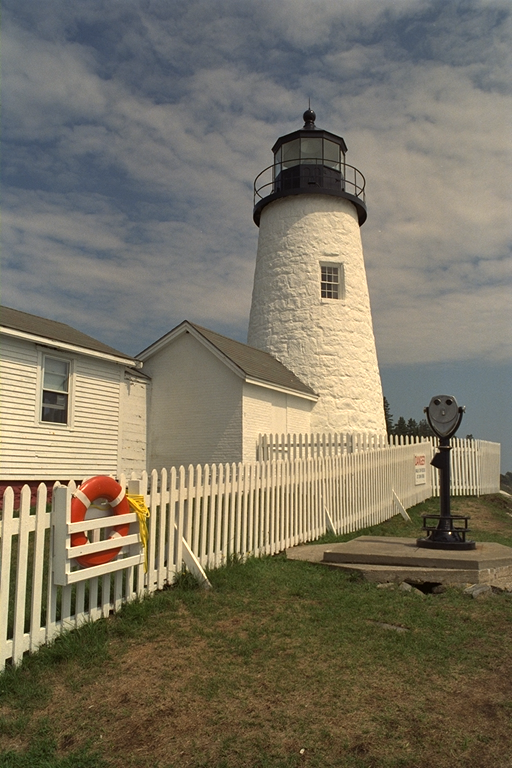}} & \multicolumn{6}{c}{\includegraphics[scale=\picsize]{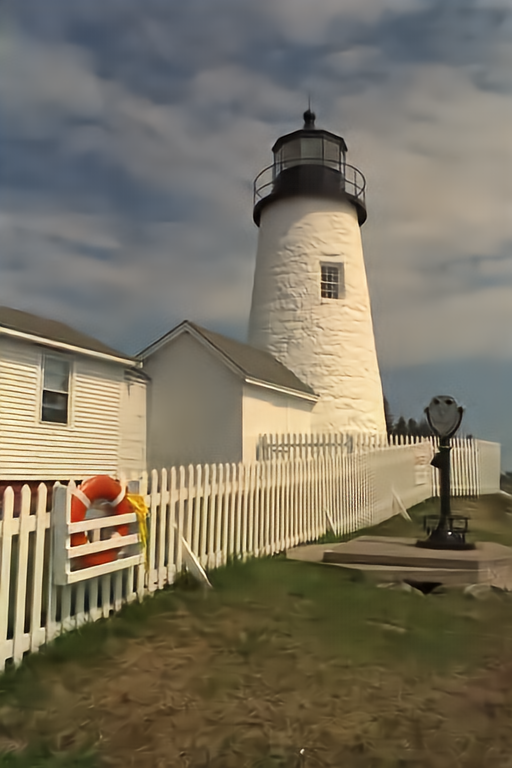}} \\
    \hline
    \hline
    \multicolumn{6}{c|}{Feature map $\hat{\mathbf{y}}_{65}$} & \multicolumn{6}{c}{Feature map $\hat{\mathbf{y}}_{51}$}\\
    \multicolumn{3}{c}{$\hat{\mathbf{y}}$} & \multicolumn{3}{c|}{$R(\hat{\mathbf{y}})$ (bits)} & \multicolumn{3}{c}{$\hat{\mathbf{y}}$} & \multicolumn{3}{c}{$R(\hat{\mathbf{y}})$ (bits)}\\
    \multicolumn{3}{c}{\includegraphics[scale=\ftsize]{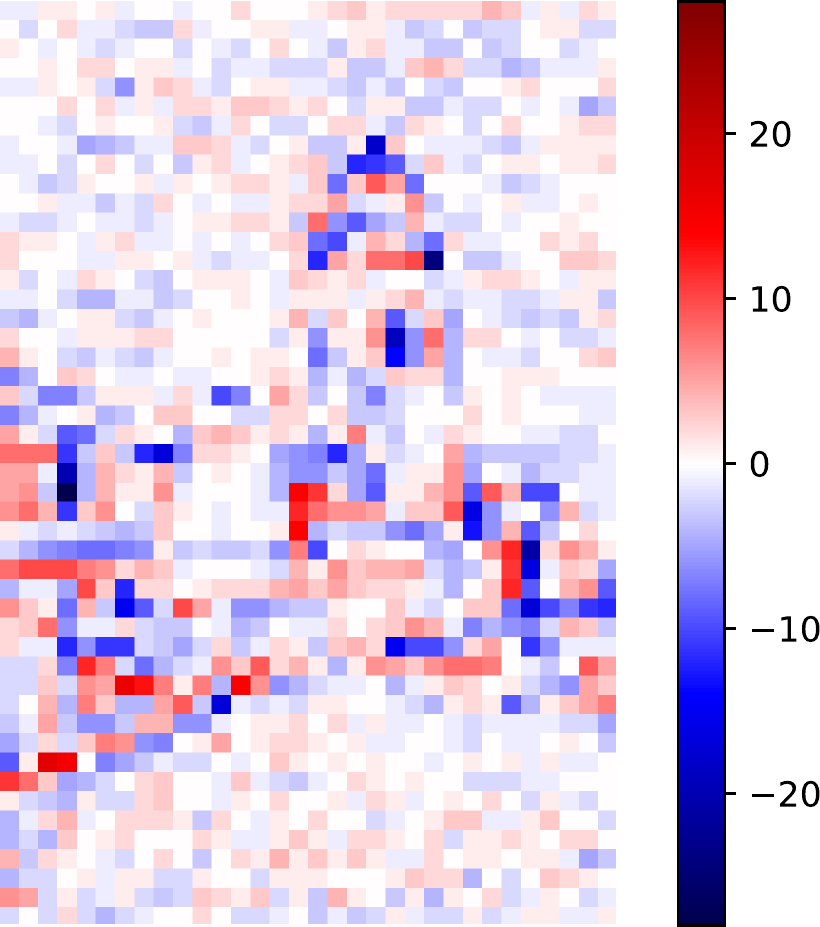}} & \multicolumn{3}{c|}{\includegraphics[scale=\ftsize]{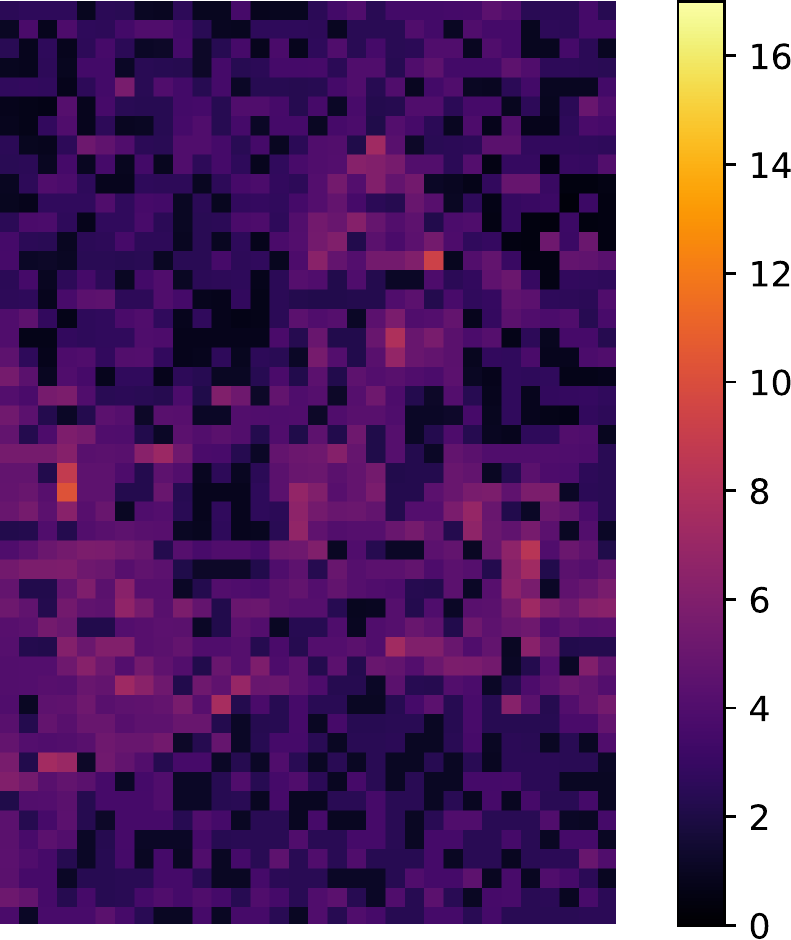}} &
    \multicolumn{3}{c}{\includegraphics[scale=\ftsize]{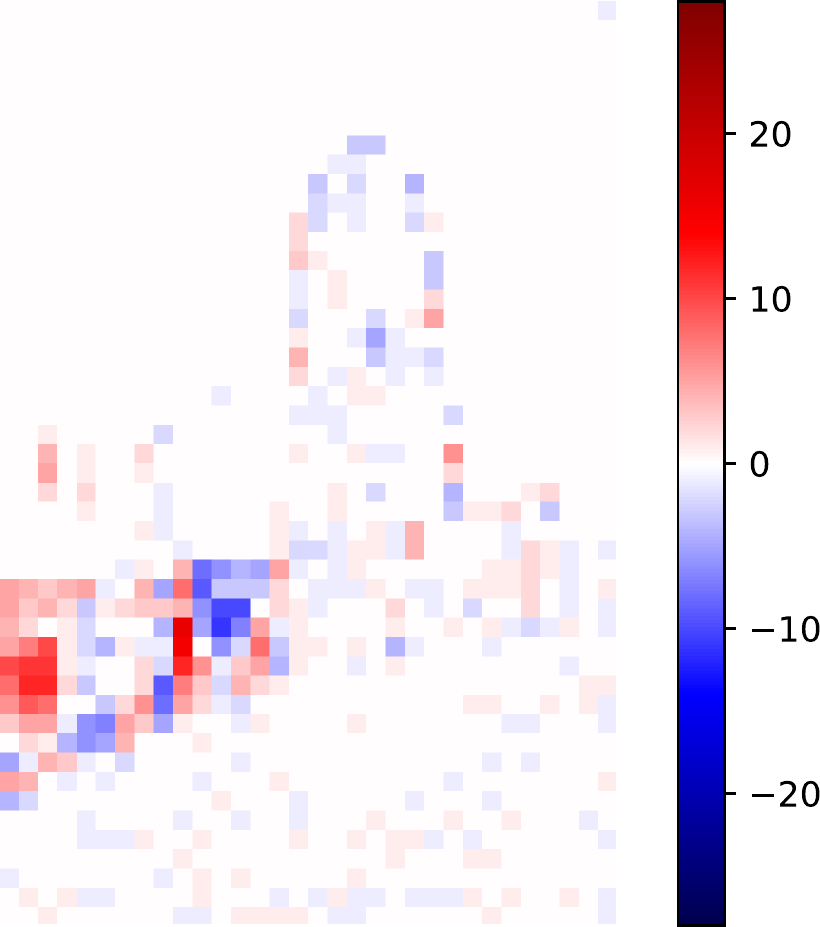}} & \multicolumn{3}{c}{\includegraphics[scale=\ftsize]{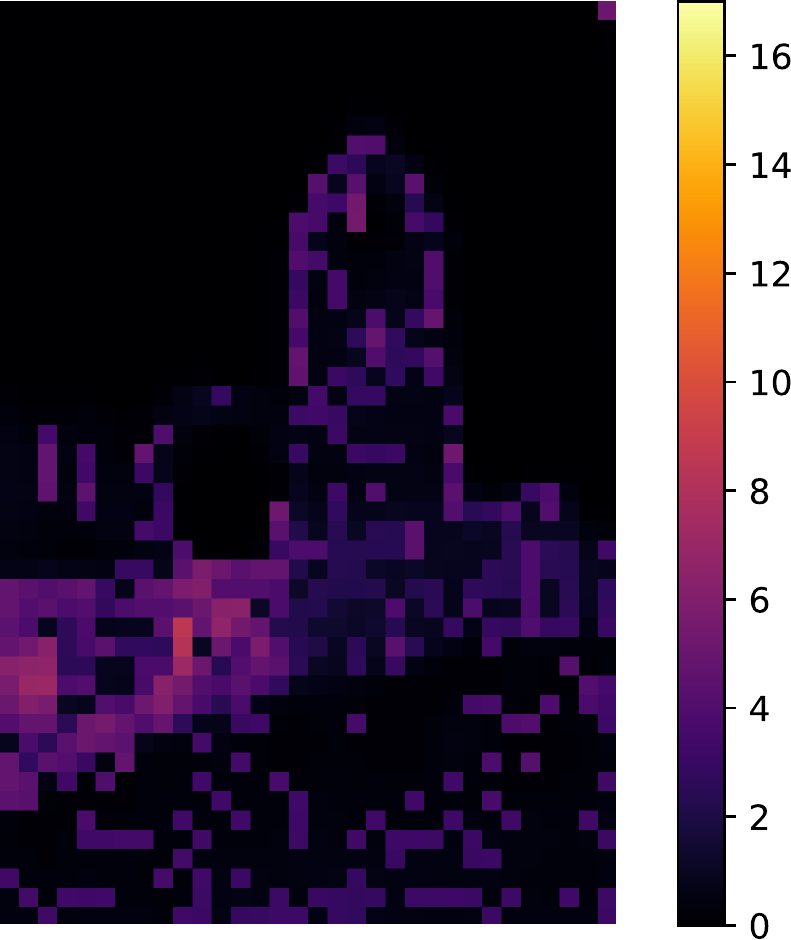}}\\
    \multicolumn{2}{c}{$\boldsymbol{P_{G_0}}$} & \multicolumn{2}{c}{$\boldsymbol{P_{G_1}}$} & \multicolumn{2}{c|}{$\boldsymbol{\sigma}$} &
    \multicolumn{2}{c}{$\boldsymbol{P_{G_0}}$} & \multicolumn{2}{c}{$\boldsymbol{P_{G_1}}$} & \multicolumn{2}{c}{$\boldsymbol{\sigma}$} \\
    \multicolumn{2}{c}{\includegraphics[scale=\mysize]{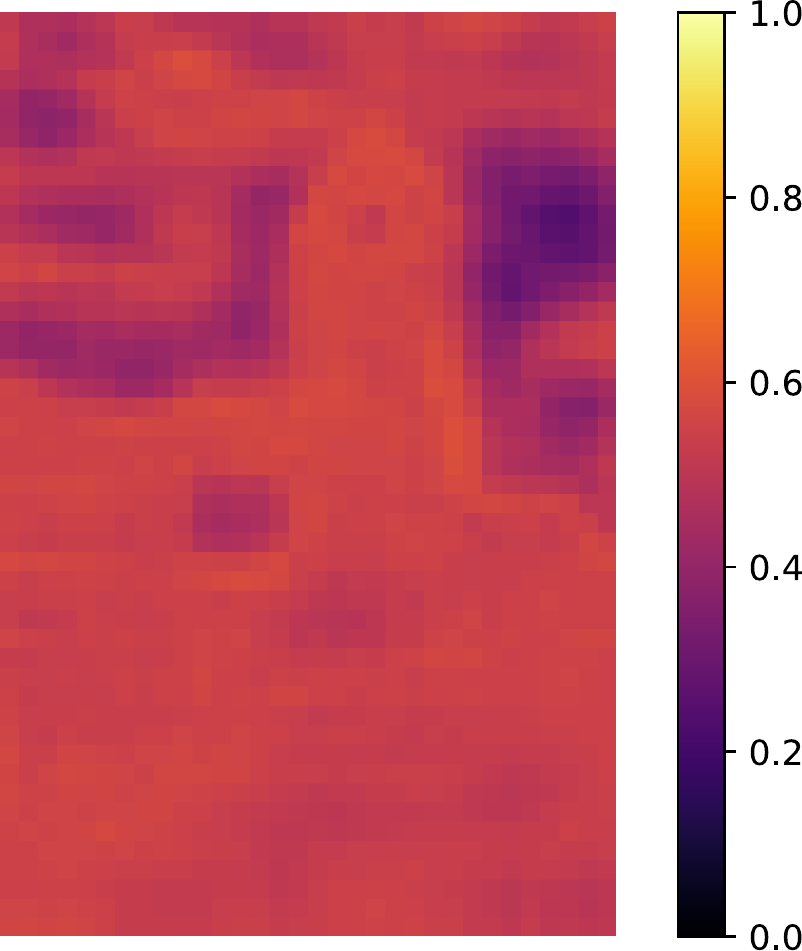}} & \multicolumn{2}{c}{\includegraphics[scale=\mysize]{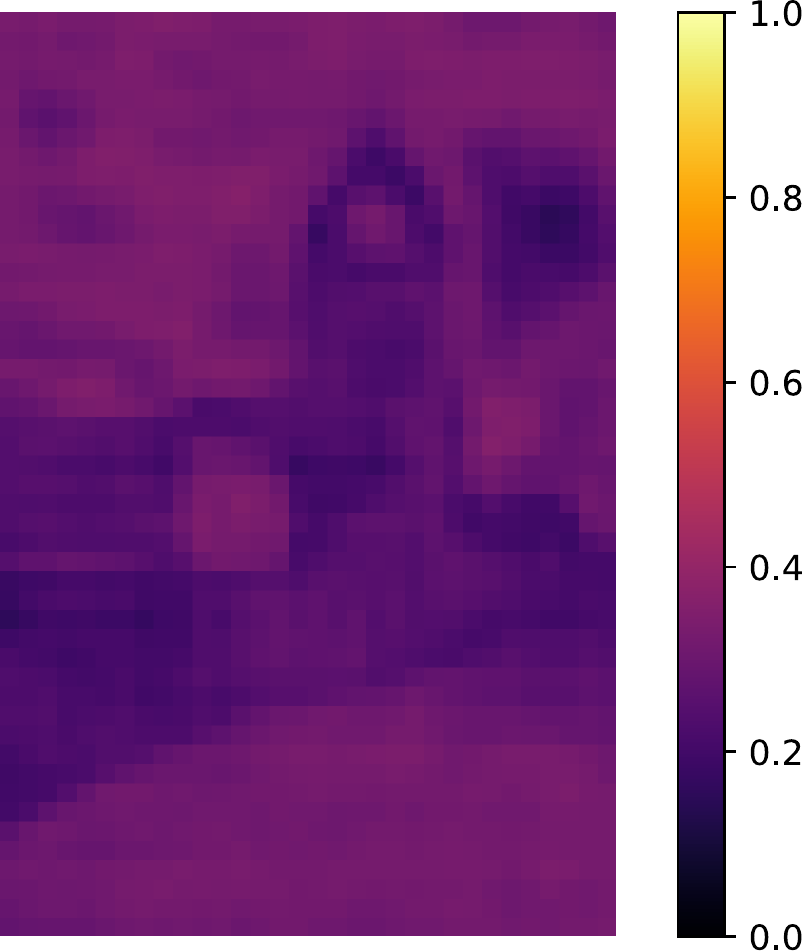}} & \multicolumn{2}{c|}{\includegraphics[scale=\mysize]{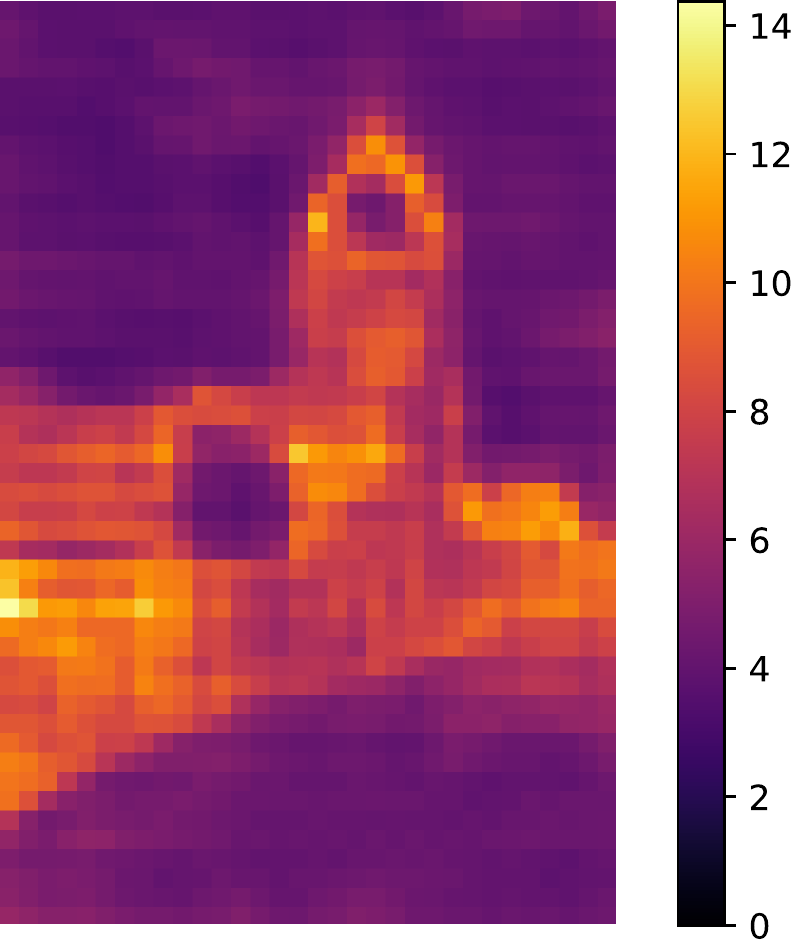}} &
    \multicolumn{2}{c}{\includegraphics[scale=\mysize]{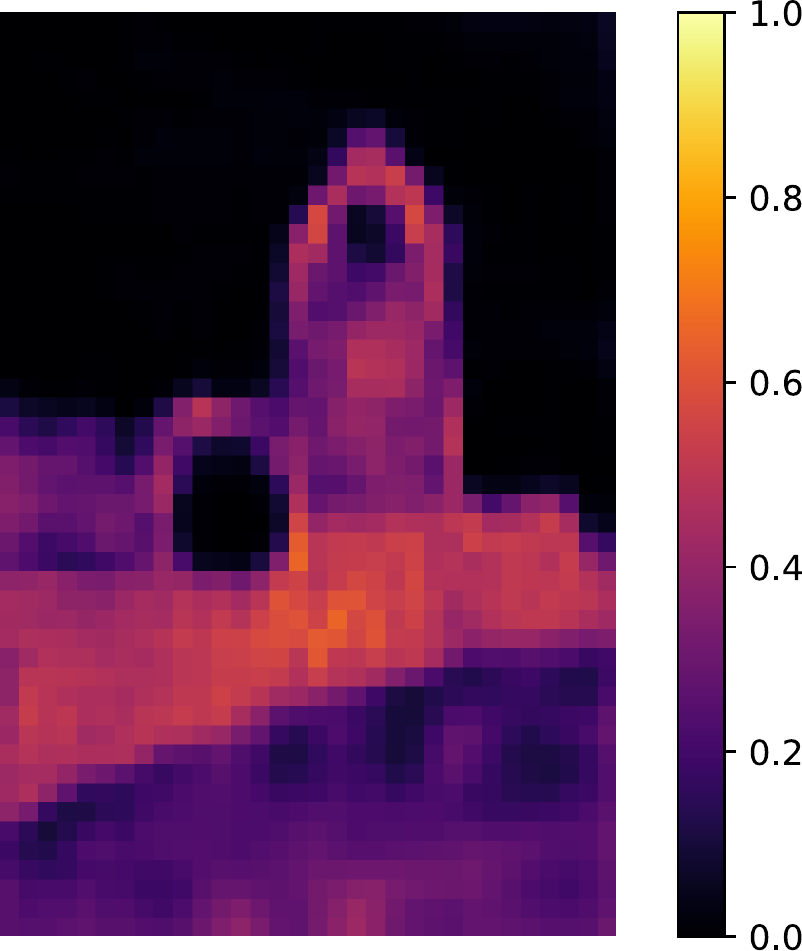}} & \multicolumn{2}{c}{\includegraphics[scale=\mysize]{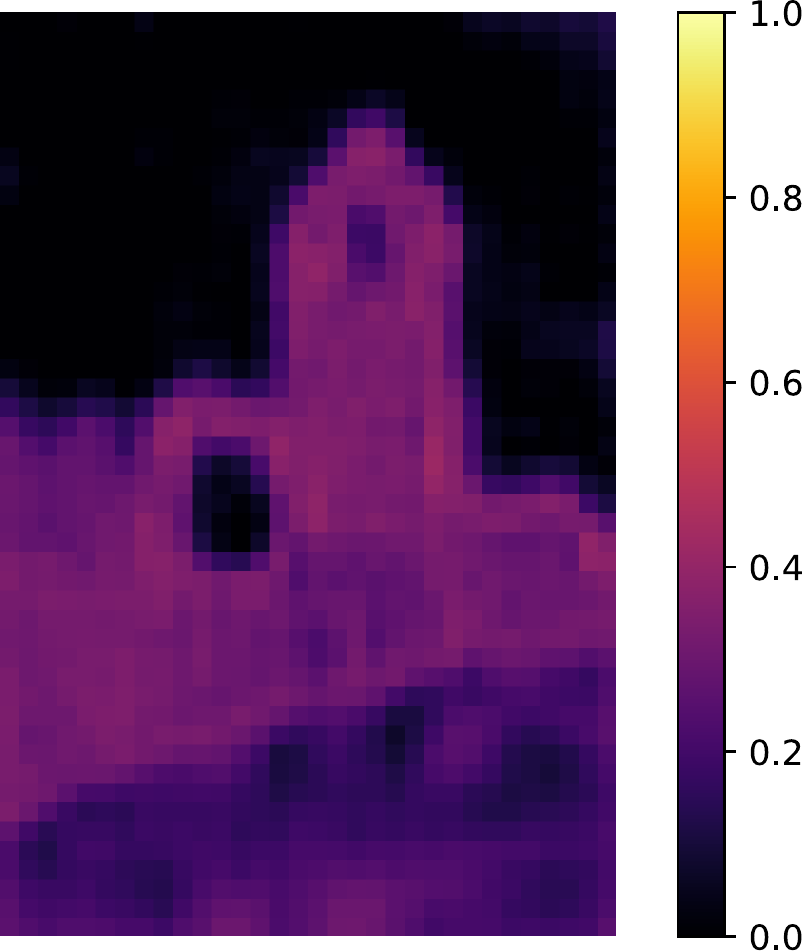}} & \multicolumn{2}{c}{\includegraphics[scale=\mysize]{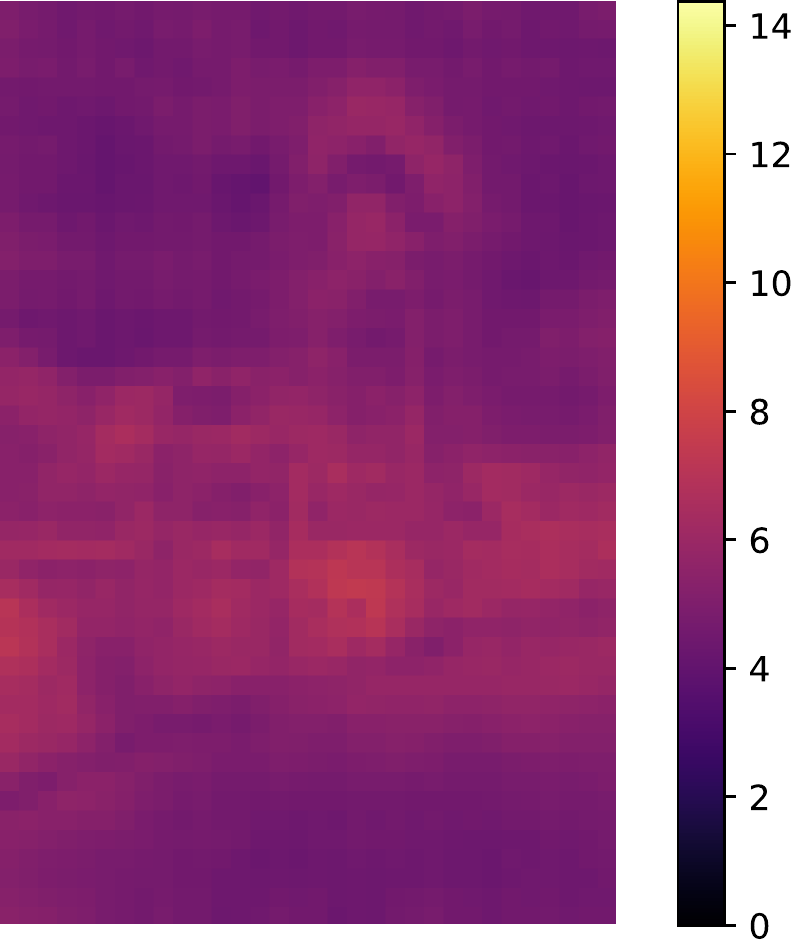}}\\
    \hline
  \end{tabular}
\end{center}
\caption{Top: Original and compressed image. Middle: two $\hat{\mathbf{y}}$ and their corresponding rate. Bottom: $\boldsymbol{P_{G_0}}$
(respectively $\boldsymbol{P_{G_1}}$) is the probability for a pixel to be
greater than 0 (respectively 1). $\boldsymbol{\sigma}$ is the scale parameter used for
explicit latents sending.}
\label{lighthouse}
\end{figure*}

The proposed method is evaluated on the CLIC 2019 low-rate task
\cite{CLIC_web_page}. The objective is to achieve the highest PSNR at 0.15 bit
per pixel (bpp). For all experiments, the training set is constructed by
concatenating the CLIC and DIV2K~\cite{Agustsson_2017_CVPR_Workshops} datasets.
The 3~000 pictures of these datasets are transformed into non-overlapping $256
\times 256$ crops. Minibatches of size 8 and Adam algorithm with a learning rate
of $10^{-4}$ are used. The training lasts 80 epochs and the learning rate is
divided by 5 at the 50\textsuperscript{th} and 70\textsuperscript{th} epoch. 

The network described in Fig. \ref{architecture} is used to evaluate three
probability models: Gaussian, Laplace and binary. For all experiments, $B=76$
$\hat{\mathbf{y}}$ features  and $C=24$ $\hat{\mathbf{z}}$ features are
transmitted.  Transforms $g_a$, $g_s$ and $h_a$ always have the same complexity.
The transform $h_s$ is slightly modified due to the number of features (denoted
as $H$ in Fig. \ref{architecture}) needed to parameterize latents distribution
($H=2B$ for Gaussian and Laplace, $H=4B$ for binary model). Hence, different
performance levels are entirely explained by the probability model. The models
are evaluated with lightweight ($M=64$) and standard ($M=192$) configurations.

The rate is estimated by the latents entropy. Performance at 0.15~bpp is obtained by
training systems with a $\lambda$ setting a working point close to the target
rate. During inference, the quantization step can be slightly deviated from 1 to
plot rate distortion curve around the training point. This enables to accurately
estimate the rate at 0.15~bpp and to compute BD rates \cite{Bjontegaard} by
comparing RD curves in [0.13, 0.17]~bpp interval. BD rate represents the rate
difference necessary to obtain the same PSNR quality between two systems.

Figure \ref{res_test} and Table \ref{comprehensive_results} sum up results on
CLIC 2019 validation and test sets, composed of 102 and 330 various resolution
images . Gaussian systems are re-implementations of Minnen et al.
\cite{DBLP:conf/nips/MinnenBT18} without the autoregressive component and are
used as a baseline. Laplacian is added as \cite{Zhou_2018_CVPR_Workshops} argues
that it slightly improves performances. BPG is also added as it is the image
version of HEVC, the state-of-the-art video coding standard.

The proposed method shows significant rate savings in all configurations, up to
18.3~\%. This proves the benefits of a richer PDF model to perform a more
efficient entropy coding. Binary probability model brings 9.1~\% rate saving for
standard systems, achieving results competitive with BPG. Performance
improvements are greater with lightweight systems. It may be because they have
less powerful transforms $g_a$ and $g_s$. Indeed, relaxing the constraints
$p_{\mathbf{y}}$ makes the system focus more on creating useful latents instead
of matching a given PDF. This holds for standard systems to a lesser extent.
Finally, it is worth noting that the binary model lightweight system can
reach the performance of the standard Gaussian system with 10 times less
parameters.

\subsection{Illustration}

Figure~\ref{lighthouse} depicts the processing of an image by the binary model
system. On the left side, feature map $\hat{\mathbf{y}}_{65}$ is the costliest
feature map (around 7~\% of the rate). Many pixels are greater than one,
resulting in high probabilities for $\boldsymbol{P_{G_0}}$ and
$\boldsymbol{P_{G_1}}$. As most of the values have important dynamic and need
explicit sending, the scale parameter $\boldsymbol{\sigma}$ takes a wide range
of values. On the right side, feature map $\hat{\mathbf{y}}_{51}$ is very sparse
and consists mostly in details, representing only 2~\% of the rate. Entirely
null areas, as the sky, are well captured by the hyperprior, with a very low
probability of being greater than zero. This allows to code them with fewer
bits.

\section{Conclusion}
This paper proposes a richer latents probability model based on binary
values and a learning process adapted for gradient-based training. Experimental
results demonstrates that this method achieves important gains compared to usual
parametric models such as Gaussian and Laplace distributions. Under the CLIC test
conditions, the binary probability model leads to a rate saving up to 18~\% for
the same reconstruction quality. In future work, the binary model could be made even more generalist
with additional flags ($G_2, G_3$ \textit{etc.}). This would reduce latents
explicit sending frequency and increase the coding performance. The
autoregressive component could be used simultaneously with the proposed binary
model to study their interactions.



\bibliographystyle{IEEEbib}
\bibliography{strings,refs}

\end{document}